\font\bbb=msbm10                                    

\overfullrule=0pt

\magnification=1200

\def\C{\hbox{\bbb C}}

\def\Z{\hbox{\bbb Z}}

\def\ForP{{\it Fortsch.\ Phys.}}

\def\MA{{\sl Math.\ Ann.}}

\def\Naw{{\sl Naturwissenschaften}}

\def\PRA{{\sl Phys.\ Rev.\ A\/}}

\def\PRL{{\sl Phys.\ Rev.\ Lett.}}
\def\PRSLA{{\sl Proc.\ Roy.\ Soc.\ Lond.\ A\/}}

\def\PhT{{\sl Phys.\ Today\/}}

\def\PTRSLA{{\sl Phil.\ Trans.\ Roy.\ Soc.\ Lond.\ A\/}}

\def\SIAMJC{{\sl SIAM J. Comput.}}

\def\SIAMR{{\sl SIAM Rev.}}
\def\SIGACTN{{\sl SIGACT News\/}}

\def\dajm{\hbox{D. A. Meyer}}

\def\vonneumann{\hbox{J. von Neumann}}
\def\schrodinger{\hbox{E. Schr\"odinger}}

\def\steane{\hbox{A. M. Steane}}

\def\hfb{\hfil\break}

\catcode`@=11
\newskip\ttglue

   \font\ninerm=cmr9    \font\eightrm=cmr8   \font\sixrm=cmr6
  \font\ninebf=cmbx9   \font\eightbf=cmbx8  \font\sixbf=cmbx6
  \font\nineit=cmti9   \font\eightit=cmti8  
  \font\ninesl=cmsl9   \font\eightsl=cmsl8  
  \font\ninemi=cmmi9   \font\eightmi=cmmi8  \font\sixmi=cmmi6

\font\bigten=cmr10 scaled\magstep2 

\def\ninepoint{\def\rm{\fam0\ninerm}%
  \textfont0=\ninerm \scriptfont0=\sixrm
  \textfont1=\ninemi \scriptfont1=\sixmi
  \textfont\itfam=\nineit  \def\it{\fam\itfam\nineit}%
  \textfont\slfam=\ninesl  \def\sl{\fam\slfam\ninesl}%
  \textfont\bffam=\ninebf  \scriptfont\bffam=\sixbf
    \def\bf{\fam\bffam\ninebf}%
  \tt \ttglue=.5em plus.25em minus.15em
  \normalbaselineskip=11pt
  \setbox\strutbox=\hbox{\vrule height8pt depth3pt width0pt}%
  \normalbaselines\rm}

\def\eightpoint{\def\rm{\fam0\eightrm}%
  \textfont0=\eightrm \scriptfont0=\sixrm
  \textfont1=\eightmi \scriptfont1=\sixmi
  \textfont\itfam=\eightit  \def\it{\fam\itfam\eightit}%
  \textfont\slfam=\eightsl  \def\sl{\fam\slfam\eightsl}%
  \textfont\bffam=\eightbf  \scriptfont\bffam=\sixbf
    \def\bf{\fam\bffam\eightbf}%
  \tt \ttglue=.5em plus.25em minus.15em
  \normalbaselineskip=9pt
  \setbox\strutbox=\hbox{\vrule height7pt depth2pt width0pt}%
  \normalbaselines\rm}

\def\sfootnote#1{\edef\@sf{\spacefactor\the\spacefactor}#1\@sf
      \insert\footins\bgroup\eightpoint
      \interlinepenalty100 \let\par=\endgraf
        \leftskip=0pt \rightskip=0pt
        \splittopskip=10pt plus 1pt minus 1pt \floatingpenalty=20000
        \parskip=0pt\smallskip\item{#1}\bgroup\strut\aftergroup\@foot\let\next}
\skip\footins=12pt plus 2pt minus 2pt
\dimen\footins=30pc

\def\ie{{\it i.e.}}
\def\eg{{\it e.g.}}
\def\etc{{\it etc.}}

\def\and{{\eightpoint AND}}
\def\not{{\eightpoint NOT}}
\def\PQpf{PQ P{\eightpoint ENNY} F{\eightpoint LIP}}
\def\GaN{G{\eightpoint UESS A} N{\eightpoint UMBER}}

\def\DeutschJozsa{1}
\def\Simon{2}
\def\Shor{3}
\def\Grover{4}
\def\Qcrypto{5}
\def\QECC{6}
\def\Qchannels{7}
\def\Qdist{8}
\def\vNcomp{9}
\def\vNqm{10}
\def\vNgame{11}
\def\vNM{12}
\def\Qstrat{13}
\def\vanEnk{14}
\def\reply{15}
\def\BBBV{16}
\def\BBHT{17}
\def\Zalka{18}
\def\Lloyd{19}
\def\Jozsa{20}
\def\Steane{21}
\def\NMRseparable{22}
\def\separabledef{23}
\def\Krauss{24}
\def\Dirac{25}
\def\EWL{26}
\def\BenjaminHayden{27}
\def\SteanevanDam{28}
\def\CEMM{29}
\def\BernsteinVazirani{30}
\def\Schrodinger{31}
\def\Schumacher{32}
\def\MeyerWallach{33}
\def\vsmodel{34}
\def\noe{35}

\input epsf.tex

\dimen0=\hsize \divide\dimen0 by 13 \dimendef\chasm=0
\dimen1=\hsize \advance\dimen1 by -\chasm \dimendef\usewidth=1
\dimen2=\usewidth \divide\dimen2 by 2 \dimendef\halfwidth=2
\dimen3=\usewidth \divide\dimen3 by 3 \dimendef\thirdwidth=3
\dimen4=\hsize \advance\dimen4 by -\halfwidth \dimendef\secondstart=4
\dimen5=\halfwidth \advance\dimen5 by -10pt \dimendef\indenthalfwidth=5
\dimen6=\thirdwidth \multiply\dimen6 by 2 \dimendef\twothirdswidth=6
\dimen7=\twothirdswidth \divide\dimen7 by 4 \dimendef\qttw=7
\dimen8=\qttw \divide\dimen8 by 4 \dimendef\qqttw=8
\dimen9=\qqttw \divide\dimen9 by 4 \dimendef\qqqttw=9
\dimen10=2.25truein \dimendef\sciencecol=10

\parskip=0pt\parindent=0pt

\line{to appear in the AMS {\sl Contemporary Mathematics\/} volume: 
                                                   \hfil 5 April 2000}
\line{{\sl Quantum Computation and Quantum Information Science} 
                                               \hfil quant-ph/0004092}
\vfill
\centerline{\bf\bigten QUANTUM GAMES AND}
\medskip
\centerline{\bf\bigten QUANTUM ALGORITHMS}
\bigskip\bigskip
\centerline{\bf David A. Meyer}
\bigskip 
\centerline{\sl Project in Geometry and Physics}
\centerline{\sl Department of Mathematics}
\centerline{\sl University of California/San Diego}
\centerline{\sl La Jolla, CA 92093-0112}
\centerline{\tt dmeyer@chonji.ucsd.edu}
\smallskip
\centerline{\sl and Institute for Physical Sciences}
\centerline{\sl Los Alamos, NM}
\smallskip

\vfill
\centerline{ABSTRACT}
\bigskip
\noindent A quantum algorithm for an oracle problem can be understood 
as a quantum strategy for a player in a two-player zero-sum game in 
which the other player is constrained to play classically.  I 
formalize this correspondence and give examples of games (and hence 
oracle problems) for which the quantum player can do better than would 
be possible classically.  The most remarkable example is the 
Bernstein-Vazirani quantum search algorithm which I show creates no 
entanglement at any timestep.

\bigskip\bigskip
\noindent 1999 Physics and Astronomy Classification Scheme:
                   03.67.Lx, 
                   02.50.Le. 

\noindent 2000 American Mathematical Society Subject Classification:
                   81P68,    
                   68Q15,    
                   91A05.    
\global\setbox1=\hbox{Key Words:\enspace}
\parindent=\wd1
\smallskip
\item{Key Words:}  quantum strategy, quantum algorithm, query 
                   complexity, entanglement.

\vfill
\hrule width2.0truein
\medskip
\noindent Expanded version of an invited talk presented at the Special
Session on Quantum Computation and Information at the Joint Mathematics
Meetings, Washington, DC, 19--22 January 2000.
\eject

\headline{\ninepoint\it Quantum games \& quantum algorithms
                                                \hfill David A. Meyer}

\parskip=10pt
\parindent=20pt

\noindent{\bf 1.  Introduction}

\noindent Despite the exuberance of people working on quantum 
computation---and more generally, quantum information theory---we 
discuss remarkably few quantum algorithms.  These include, and are
largely limited to, the Deutsch-Jozsa [\DeutschJozsa], Simon [\Simon],
Shor [\Shor] and Grover [\Grover] algorithms.  Of course, quantum 
versions of many other information processing tasks have also been
studied:  cryptography [\Qcrypto], error correction [\QECC], 
communication channels [\Qchannels], distributed computation [\Qdist],
\etc\ \ When I was invited to speak about quantum computing at 
Microsoft Research in January 1998, I decided to try to add game 
theory to this list.

My motivations were twofold:  First, as I explained in that talk,
von~Neumann was not only the driving force behind the development of 
modern digital computers [\vNcomp]---a subject of great interest to 
Microsoft---but also one of the founders of quantum mechanics [\vNqm],
and thus someone whose ideas are central to quantum computing.  But
he had another great interest---shared with Microsoft---economics!  
Von~Neumann essentially invented game theory [\vNgame]; his book with
Morgenstern, {\sl Theory of Games and Economic Behavior\/} [\vNM]
raises (and in some cases, answers) many of the questions which 
preoccupy game theorists and economists today.  Second, I hoped that 
something like the argument that identifies which two-person zero-sum
games have optimal mixed, rather than pure, classical strategies might
provide some insight into which problems are solvable more efficiently
by quantum rather than classical algorithms.  This hope was probably
somewhat na{\"\i}ve, but it brings us to the first question I'll 
address in this talk:  What do quantum games have to do with quantum
algorithms?

The quantum game I described originally [\Qstrat], \PQpf, is perhaps 
too simple to make the connection with quantum algorithms completely 
clear.  In fact, it is so simple, involving only one qubit, that 
several people have pointed out that it could be simulated classically 
[\vanEnk].  I have argued elsewhere that this misses the point 
slightly, that the issue is not whether there exists a classical 
simulation, but how the complexity of that simulation would scale if 
the size of the game were to increase [\reply].  To illustrate this 
point explicitly, in this talk I'll tell a story which involves a game 
which has instances of arbitrarily large size.  My discussion of this 
game naturally includes a description of Grover's algorithm [\Grover], 
which not only helps to answer the first question, but raises a second 
and third.

The second is:  Are there sophisticated quantum search algorithms?
More explicitly, are there `databases' which can be `searched' with
better than the square root speedup that Grover's algorithm provides
over the best possible classical algorithm?~[\Grover]\ \ Since 
Bennett, Bernstein, Brassard \& Vazirani [\BBBV], Boyer, Brassard, 
H{\o}yer and Tapp [\BBHT], and Zalka [\Zalka] have shown that Grover's 
algorithm is optimal, I will explain the natural changes in the 
problem which make this question interesting.

Recently Lloyd has argued that Grover's algorithm can be implemented
without entanglement [\Lloyd].  At first glance this may appear 
surprising:  many people have stated that the power of quantum 
computing derives from entanglement [\Jozsa,\Steane].  This belief 
underlies the criticism that NMR experiments do not realize quantum 
computation because the state of the system at each timestep is 
separable [\NMRseparable], \ie, a convex combination of unentangled 
pure states [\separabledef].  Similarly, van~Enk observed that the one 
qubit \PQpf\ game not only can be simulated classically, but involves 
no entanglement [\vanEnk], which would suggest by the same `reasoning' 
that such a quantum game would be unrelated to quantum algorithms.  In 
fact, the quantum game in the story to come provides an answer to the 
third question I'll address:  Can quantum-over-classical improvements 
be achieved without entanglement? 

\medskip
\noindent{\bf 2.  PQ games}

\noindent I'll begin by reviewing briefly the one qubit game \PQpf\
[\Qstrat].  In our first episode, the starship Enterprise is facing 
some imminent catastrophe when the superpowerful being Q appears on 
the bridge and offers to rescue the ship if Captain Picard%
\sfootnote{$^*$}{Captain Picard and Q are characters in the popular 
American television (and movie) series {\sl Star Trek:  The Next 
Generation\/} whose initials and abilities are ideal for this
illustration.  See [\Krauss].}
can beat him at a simple game:  Q produces a penny and asks Picard to
place it in a small box, head up.  Then Q, followed by Picard, 
followed by Q, reaches into the box, without looking at the penny, and
either flips it over or leaves it as it is.  After Q's second turn 
they open the box and Q wins if the penny is head up.  Q wins every 
time they play, using the following quantum strategy:
$$
\global\setbox2=\hbox{$\scriptstyle\sigma_x{\rm\ or\ }I_2$}
\eqalign{
|0\rangle
 &\mathrel{\mathop{\kern0pt\longrightarrow}\limits^{\rm 
  Q}_{\hbox to\wd2{\hfil $\scriptstyle H$\hfil}}}
{1\over\sqrt{2}}(|0\rangle + |1\rangle)                            \cr
 &\mathrel{\mathop{\kern0pt\longrightarrow}\limits^{\rm 
  Picard}_{\sigma_x{\rm\ or\ }I_2}}
{1\over\sqrt{2}}(|0\rangle + |1\rangle)                            \cr
 &\mathrel{\mathop{\kern0pt\longrightarrow}\limits^{\rm 
  Q}_{\hbox to\wd2{\hfil $\scriptstyle H$\hfil}}}
|0\rangle                                                          \cr
}
$$
Here $|\cdot\rangle$ is Dirac notation [\Dirac] for an element of 
Hilbert space, 0 denotes `head' and 1 denotes `tail',
$H = {1\over\sqrt{2}}\bigl({1 \atop 1} {1 \atop -1}\bigr)$ is the 
Hadamard transformation, and 
$\sigma_x = \bigl({0 \atop 1} {1 \atop 0}\bigr)$ implements Picard's 
possible action of flipping the penny over.  Q's quantum strategy of 
putting the penny into the equal superposition of `head' and `tail' on 
his first turn means that whether Picard flips the penny over or not, 
it remains in an equal superposition which Q can rotate back to `head' 
by applying $H$ again since $H = H^{-1}$.  So when they open the box, 
Q always wins.

Notice that if Q were restricted to playing classically, \ie, to 
implementing only $\sigma_x$ or $I_2$ on his turns, an optimal 
strategy for both players would be to flip the penny over or not with 
equal probability on each turn.  In this case Q would win only half 
the time, so he does substantially better by playing quantum 
mechanically.

The structure of \PQpf\ motivates the following definition which 
formalizes one meaning for quantum game.%
\sfootnote{$^{\dagger}$}{Eisert, Wilkens \& Lewenstein have proposed a
different formalism for quantum games and have applied it to find a 
(unique) Pareto optimal equilibrium for the Prisoners' Dilemma when 
the players are allowed to select moves from the same specific subset 
of unitary transformations [\EWL].  Benjamin \& Hayden recently showed 
that this is not a solution when the players are allowed to make 
arbitrary unitary moves [\BenjaminHayden]; in fact, in this case there 
is no equilibrium, just as in PQ P{\sixrm ENNY} F{\sixrm LIP} 
[\Qstrat].  Nevertheless, that formalism still seems likely to be 
interesting, although more closely related to quantum communication 
protocols [\Qdist,\SteanevanDam] than to quantum algorithms.}
%

\noindent{\bf Definitions}.  A {\sl PQ game\/} consists of

\itemitem{({\it i\/})} a Hilbert space ${\cal H}$---the possible 
states of the game---with $N = {\rm dim} {\cal H}$,

\itemitem{({\it ii\/})} an initial state $\psi_0 \in {\cal H}$,

\itemitem{({\it iii\/})} subsets $Q_i \subset U(N)$, 
$i\in\{1,\ldots,k+1\}$---the elements of $Q_i$ are the moves Q chooses
among on turn $i$,

\itemitem{({\it iv\/})} subsets $P_i \subset S_N$, 
$i\in\{1,\ldots,k\}$, where $S_N$ is the permutation group on $N$ 
elements---the elements of $P_i$ are the moves Picard chooses among on 
turn $i$, and

\itemitem{({\it v\/})} a projection operator $\Pi$ on ${\cal H}$---the
subspace $W_{\rm Q}$ fixed by $\Pi$ consists of the winning states for
Q.

Since only Picard and Q play, these are {\sl two-player\/} games; they
are {\sl zero-sum\/} since when Q wins, Picard loses, and {\it vice
versa}.

A {\sl pure quantum strategy\/} for Q is a sequence $u_i \in Q_i$.  A
{\sl pure\/} ({\sl classical\/}) {\sl strategy\/} for Picard is a 
sequence $s_i \in P_i$, while a {\sl mixed\/} ({\sl classical\/}) 
{\sl strategy\/} for Picard is a sequence of probability distributions 
$f_i : P_i \to [0,1]$.  If both Q and Picard play pure strategies, the 
corresponding {\sl evolution\/} of the PQ game is described by
$$
\psi_f = u_{k+1} s_k u_k \ldots u_2 s_1 u_1 \psi_0.
$$
After Q's last move the state of the game is measured with $\Pi$.
According to the rules of quantum mechanics [\vNqm], the players
observe the eigenvalue 1 with probability 
${\rm Tr}(\psi^{\dagger}\Pi\psi)$; this is the probability that the 
state is projected into $W_{\rm Q}$ and Q wins.  More generally, if 
Picard plays a mixed strategy, the corresponding {\sl evolution\/} of 
the PQ game is described by
$$
\rho_f^{\vphantom\dagger}
 = 
u_{k+1}^{\vphantom\dagger}
 \Bigl(\sum_{s_k\in P_k} f_k^{\vphantom\dagger}
                        (s_k^{\vphantom\dagger}) 
                         s_k^{\vphantom\dagger}
  u_k^{\vphantom\dagger}
   \ldots
    u_2^{\vphantom\dagger}
     \Bigl(\sum_{s_1\in P_1} f_1^{\vphantom\dagger}
                            (s_1^{\vphantom\dagger}) 
                             s_1^{\vphantom\dagger}
      u_1^{\vphantom\dagger}
       \rho_0^{\vphantom\dagger}
      u_1^{\dagger}
     s_1^{\dagger}\Bigr)
    u_2^{\dagger}
   \ldots
  u_k^{\dagger}
 s_k^{\dagger}\Bigr)
u_{k+1}^{\dagger},
$$
where $\rho_0^{\vphantom\dagger} = \psi_0^{\vphantom\dagger} \otimes 
                                   \psi_0^{\dagger}$.  
Again, after Q's last move $\rho_f$ is measured with $\Pi$; the 
probability that $\rho_f$ is projected into 
$W_{\rm Q}^{\vphantom\dagger} \otimes W_{\rm Q}^{\dagger}$) and Q wins
is ${\rm Tr}(\Pi\rho_f)$.

Finally, an {\sl equilibrium\/} is a pair of strategies, one for 
Picard and one for Q, such that neither player can improve his 
probability of winning by changing his strategy while the other does 
not.

As I'll show in the next section, the structure of a PQ game 
specializes to the structure of the known quantum algorithms.  In 
general, unlike the simple case of \PQpf, 
$W_{\rm Q} = W_{\rm Q}(\{s_i\})$ or $W_{\rm Q} = W_{\rm Q}(\{f_i\})$, 
\ie, the conditions for Q's win can depend on Picard's strategy.  Each
of the three example games I consider here suffices to prove that there
are games with mixed/quantum equilibria at which Q does better than he 
would at any mixed/mixed equilibrium; equivalently, there are some 
quantum algorithms which outperform classical ones.

\medskip
\noindent{\bf 3.  Guessing a number}

\noindent In our second episode, Q returns to the Enterprise and
challenges Captain Picard again.  He boasts that if Picard picks any
number between 0 and $N-1$, inclusive, he can guess it.  Now, Picard
is no slouch; he has been studying up on quantum algorithms since the 
last episode.  In particular, he has studied Grover's algorithm 
[\Grover] and realizes that for $N = 2^n$, Q can determine the number 
he picks with high probability by playing the following strategy:
$$
\global\setbox2=\hbox{$\scriptstyle H^{\otimes n}\otimes I_2 %
                            \circ s(f_0) %
                            \circ H^{\otimes n}\otimes I_2$}
\global\setbox3=\hbox{$u_1$}
\eqalignno{
|0\ldots0,0\rangle
 &\mathrel{\mathop{\kern0pt\longrightarrow}\limits^{\rm 
   Q}_{\hbox to\wd2{\hfil $\scriptstyle 
       H^{\otimes n}\otimes H\sigma_x$\hfil}}}
  {1\over\sqrt{N}}\sum_{x=0}^{N-1}
   |x\rangle\otimes{1\over\sqrt{2}}(|0\rangle - |1\rangle)   &(u_1)\cr
 &\mathrel{\mathop{\kern0pt\longrightarrow}\limits^{\rm 
   Picard}_{\hbox to\wd2{\hfil $\scriptstyle s(f_a)$\hfil}}}
  {1\over\sqrt{N}}\sum_{x=0}^{N-1} (-1)^{\delta_{xa}}
   |x\rangle\otimes{1\over\sqrt{2}}(|0\rangle - |1\rangle)   
                                             &(\hbox to\wd3{$s_1$})\cr
 &\mathrel{\mathop{\kern0pt\longrightarrow}\limits^{\rm 
   Q}_{H^{\otimes n}\otimes I_2 \circ s(f_0) 
   \circ H^{\otimes n}\otimes I_2}}
  \cdots,                                                    &(u_2)\cr
}
$$
where $a \in [0,N-1]$ is Picard's chosen number, and the moves $s_1$ 
and $u_2$ are repeated a total of 
$k = \lfloor{\pi\over4}\sqrt{N}\rfloor$ times, \ie, 
$s_k = \cdots = s_1$ and $u_{k+1} = \cdots = u_2$.  For 
$f : \Z_2^n \to \Z_2$, $s(f)$ is the permutation (and hence unitary 
transformation) defined by
$$
s(f)|x,b\rangle = |x,b\oplus f(x)\rangle,
$$
where $\oplus$ denotes addition mod 2.  This transformation is often 
referred to as `$f$-controlled-\not'.\ \ Each of Picard's moves $s_i$ 
can be thought of as the response of an oracle which computes 
$f_a(x) := \delta_{xa}$ to respond to the quantum query defined by the 
state after $u_i$.  After $O(\sqrt{N})$ such queries, a measurement by 
$\Pi = |a\rangle\langle a| \otimes I_2$ returns a win for Q with 
probability bounded above ${1\over2}$, \ie, Grover's quantum algorithm
determines $a$ with high probability.  (Here 
$\langle a| := |a\rangle^{\dagger}$ and the tensor product is implicit
in $|a\rangle\langle a|$ [\Dirac].)

Notice that if Q were to play classically, he could query Picard about
a specific number at each turn, but on the average it would take $N/2$
turns to guess $a$.  A classical equilibrium is for Picard to choose 
$a$ uniformly at random, and for Q to choose a permutation of $N$ 
uniformly at random and guess numbers in the corresponding order.  
Even when Picard plays such a mixed strategy, Q's quantum strategy is 
optimal; together they define a mixed/quantum equilibrium.

Knowing all this, Picard responds that he will be happy to play, but
that Q should only get 1 guess, not 
$\lfloor{\pi\over4}\sqrt{N}\rfloor$.  Q protests that this is hardly
fair, but he will play, as long as Picard tells him how close his 
guess is to the chosen number.  Picard agrees, and they play.  Q wins.
They play again.  Q wins again.  Picard doesn't understand what's
going on.  The problem is that in his studies of quantum algorithms, 
he overlooked an insufficiently appreciated quantum algorithm, the
slightly improved [\CEMM] Bernstein-Vazirani algorithm 
[\BernsteinVazirani]:  Guess $x$ and answer $a$ are vectors in 
$\Z_2^n$, so $x\cdot a$ depends on the cosine of the angle between the 
vectors.  Thus it seems reasonable to define ``how close a guess is to
the answer'' to be the oracle response $g_a(x) := x\cdot a$.  Then Q
plays as follows:
$$
\global\setbox2=\hbox{$\scriptstyle H^{\otimes n}\otimes H\sigma_x$}
\global\setbox3=\hbox{$u_1$}
\eqalignno{
|0\ldots0,0\rangle
 &\mathrel{\mathop{\kern0pt\longrightarrow}\limits^{\rm 
   Q}_{H^{\otimes n}\otimes H\sigma_x}}
  {1\over\sqrt{N}}\sum_{x=0}^{N-1}
   |x\rangle\otimes{1\over\sqrt{2}}(|0\rangle - |1\rangle)   &(u_1)\cr
 &\mathrel{\mathop{\kern0pt\longrightarrow}\limits^{\rm 
   Picard}_{\hbox to\wd2{\hfil $\scriptstyle s(g_a)$\hfil}}}
  {1\over\sqrt{N}}\sum_{x=0}^{N-1} (-1)^{x\cdot a}
   |x\rangle\otimes{1\over\sqrt{2}}(|0\rangle - |1\rangle)
                                             &(\hbox to\wd3{$s_1$})\cr
 &\mathrel{\mathop{\kern0pt\longrightarrow}\limits^{\rm 
   Q}_{\hbox to\wd2{\hfil $\scriptstyle 
       H^{\otimes n}\otimes I_2$\hfil}}}
  |a\rangle\otimes{1\over\sqrt{2}}(|0\rangle - |1\rangle)    &(u_2)\cr
}
$$
For $\Pi = |a\rangle\langle a| \otimes I_2$ again, Q wins with 
probability 1, having queried Picard only once!

Just as before, Picard makes the game hardest for Q classically if he
chooses $a$ uniformly at random.  Classically Q requires $n$ queries
to determine $a$ with probability 1.  The classical to quantum 
improvement in number of queries is thus $n$ to 1, in some sense 
greater than the Grover improvement from $O(N)$ to $O(\sqrt{N})$.

\medskip
\noindent{\bf 4.  Entanglement}

\noindent Most remarkably, Bernstein \& Vazirani's `sophisticated' 
quantum search algorithm achieves this improvement without creating 
any entanglement at any timestep!  Recall the following definition.

\noindent{\bf Definition}.  A {\sl pure state\/}---a vector in 
${\cal H}$---is {\sl entangled\/} if it does not factor relative to a
given tensor product decomposition of the Hilbert space 
[\Schrodinger].

In the two algorithms considered in the previous sections, the Hilbert 
space decomposes into a tensor product of {\sl qubits\/} 
[\Schumacher], \ie, $\C^2$s.  To see that the slightly improved 
[\CEMM] Bernstein-Vazirani algorithm [\BernsteinVazirani] creates no 
entanglement relative to this decomposition, note that $\psi_0$ has no 
entanglement, and hence $u_1 \psi_0$ has none since $u_1$ is the 
tensor product of operations on individual qubits.  Also, $\psi_f$ is
not entangled since $|a\rangle$ is just the tensor product of qubits 
in states $|0\rangle$ or $|1\rangle$.  But $\psi_f$ is obtained from
the intermediate state $s_1 u_1 \psi_0$ by the action of $u_2$ which,
like $u_1$, is the tensor product of operations on individual qubits. 
So $s_1 u_1 \psi_0$ also is not entangled.

In contrast, for Grover's algorithm, every state after $u_1 \psi_0$ is
entangled for $n > 1$.  By a natural measure the entanglement 
oscillates with period ${\pi\over4}\sqrt{N}$, \ie, after 
$\lfloor{\pi\over4}\sqrt{N}\rfloor$ queries the state is close to 
$|a\rangle\otimes{1\over\sqrt{2}}(|0\rangle - |1\rangle)$ and its 
entanglement is fairly small, while after only 
$\lfloor{\pi\over8}\sqrt{N}\rfloor$ queries the entanglement is fairly 
large [\MeyerWallach].  So what could be the suggestion of Lloyd, to
which I alluded in the Introduction, that Grover's algorithm can be 
implemented without entanglement [\Lloyd]?  He simply observes that 
since the first $n$ qubits are never entangled with the last qubit, if 
they are implemented by a single tensor factor of dimension $N$, there 
is no entanglement.  This is true, of course, by definition, and has 
been observed earlier in the general quantum computing setting by 
Jozsa and Ekert [\Jozsa].  The cost incurred for realizing such a 
scheme physically, as Lloyd acknowledges, increases exponentially with 
$n$ and must be paid with increasing energy, mass, or precision.

\medskip
\noindent{\bf 5.  Conclusion}

\noindent So what do quantum games have to do with quantum algorithms?
The two versions of \GaN\ illustrate the relation for oracle problems.
In these cases, quantum algorithms specialize the PQ game definition 
to require that $s_1 = \cdots = s_k$, \ie, the oracle always responds 
the same way once a function has been chosen.  Furthermore, the 
winning states for Q, $W_{\rm Q}$, depend on Picard's strategy 
$\{s_i\}$, \ie, on $a$.  The Deutsch-Jozsa [\DeutschJozsa], Simon 
[\Simon] and Shor [\Shor] algorithms can also be described this way.

Are there `sophisticated' quantum algorithms?  Yes.  The oracle which
responds in the Bernstein-Vazirani scenario with $x\cdot a$ mod 2 is a
`sophisticated database' by comparison with Grover's `na{\"\i}ve
database' which only responds that a guess is correct or incorrect. 
The former is closely related to the vector space model for 
information retrieval in which there is a vector space with basis 
vectors corresponding to the occurrence of key words:  database 
elements define vectors in this space and are ranking according to
their inner product with the vector representing a query [\vsmodel].  
Furthermore, relatively speaking, it improves more over the classical 
optimum than does Grover's algorithm.

And finally, is entanglement required for quantum-over-classical
improvements?  No.  I've shown that, remarkably, the slightly improved
version of the Bernstein-Vazirani algorithm does not create 
entanglement at any timestep, but still solves this oracle problem 
with fewer queries than is possible classically.  Relative to the 
oracle, this quantum algorithm has no entanglement, unlike Grover's, 
which does---at least within the standard model of quantum computing.  
It illustrates both `sophisticated quantum search' without 
entanglement and sophisticated ``quantum search without entanglement''
[\noe].

\medskip
\noindent{\bf Acknowledgements}
\nobreak

\nobreak
\noindent I thank Mike Freedman, Manny Knill, Raymond Laflamme, John 
Smolin and Nolan Wallach for useful discussions.  This work has been 
partially supported by Microsoft Research and by ARO grant 
DAAG55-98-1-0376.

\vfill\eject
\medskip
\global\setbox1=\hbox{[00]\enspace}
\parindent=\wd1

\noindent{\bf References}
\bigskip

\parskip=0pt
\item{[\DeutschJozsa]}
D. Deutsch \& R. Jozsa,
``Rapid solution of problems by quantum computation'',
\PRSLA\ {\bf 439} (1992) 553--558.

\item{[\Simon]}
D. R. Simon,
``On the power of quantum computation'',
in S. Goldwasser, ed.,
{\sl Proceedings of the 35th Symposium on Foundations of Computer 
Science}, Santa Fe, NM, 20--22 November 1994
(Los Alamitos, CA:  IEEE Computer Society Press 1994) 116--123.

\item{[\Shor]}
P. W. Shor,
``Algorithms for quantum computation:  discrete logarithms and 
  factoring'',
in S. Goldwasser, ed.,
{\sl Proceedings of the 35th Symposium on Foundations of Computer 
Science}, Santa Fe, NM, 20--22 November 1994
(Los Alamitos, CA:  IEEE Computer Society Press 1994) 124--134.

\item{[\Grover]}
L. K. Grover,
``A fast quantum mechanical algorithm for database search'',
in 
{\sl Proceedings of the 28th Annual ACM Symposium on the Theory of
Computing}, Philadelphia, PA, 22--24 May 1996
(New York:  ACM 1996) 212--219.

\item{[\Qcrypto]}
S. Wiesner, 
``Conjugate coding'',
\SIGACTN\ {\bf 15} (1983) 78--88;\hfb
C. H. Bennett \& G. Brassard,
``Quantum cryptography:  Public-key distribution and coin tossing'',
in 
{\sl Proceedings of the IEEE International Conference on Computers,
Systems and Signal Processing}, Bangalore, India, December 1984
(New York:  IEEE 1984) 175--179;\hfb
A. Ekert,
``Quantum cryptography based on Bell's theorem'',
\PRL\ {\bf 67} (1991) 661--663.

\item{[\QECC]}
P. W. Shor,
``Scheme for reducing decoherence in quantum computer memory'',
\PRA\ {\bf 52} (1995) R2493--R2496;\hfb
A. M. Steane,
``Error correcting codes in quantum theory'',
\PRL\ {\bf 77} (1996) 793--797.

\item{[\Qchannels]}
A. S. Kholevo, 
``Bounds for the quantity of information transmitted by a quantum
  communication channel'',
{\it Problemy Peredachi Informatsii\/} {\bf 9} (1973) 3--11; 
transl.\ in
{\sl Problems Inf.\ Transmiss.}\ {\bf 9} (1973) 177--183;\hfb
C. H. Bennett, D. P. DiVincenzo, J. Smolin \& W. K. Wootters,
``Mixed state entanglement and quantum error correction'',
\PRA\ {\bf 54} (1996) 3824--3851;\hfb
B. Schumacher, M. Westmoreland \& W. K. Wootters,
``Limitation on the amount of accessible information in a quantum
  channel'',
\PRL\ {\bf 76} (1997) 3452--3455.

\item{[\Qdist]}
R. Cleve \& H. Buhrman,
``Substituting quantum entanglement for communication'',
\PRA\ {\bf 56} (1997) 1201--1204;\hfb
H. Buhrman, W. van~Dam, P. H{\o}yer \& A. Tapp,
``Multiparty quantum communication complexity'',
\PRA\ {\bf 60} (1999) 2737--2741.

\item{[\vNcomp]}
\vonneumann,
in 
A. H. Taub, ed.,
{\sl Collected Works},
Vol.\ 5, 
{\sl Design of Computers, Theory of Automata and Numerical 
  Analysis\/}
(New York: Pergamon Press 1961--1963).

\item{[\vNqm]}
\vonneumann,
{\it Mathematische Grundlagen der Quantenmechanik\/}
(Berlin:  Spring\-er-Verlag 1932); 
transl.\ by R. T. Beyer as
{\sl Mathematical Foundations of Quantum Mechanics\/}
(Princeton:  Princeton University Press 1955).

\item{[\vNgame]}
\vonneumann,
``{\it Zur theorie der gesellschaftsspiele\/}'',
\MA\ {\bf 100} (1928) 295--320.

\item{[\vNM]}
\vonneumann\ \& O. Morgenstern,
{\sl Theory of Games and Economic Behavior}, third edition
(Princeton:  Princeton University Press 1953).

\item{[\Qstrat]}
\dajm,
``Quantum strategies'',
\PRL\ {\bf 82} (1999) 1052--1055.

\item{[\vanEnk]}
See, \eg, 
S. J. van Enk,
``Quantum and classical game strategies'',
\PRL\ {\bf 84} (2000) 789.

\item{[\reply]}
\dajm,
``Why quantum strategies are quantum mechanical'',
published as\break
``Meyer replies'',
\PRL\ {\bf 84} (2000) 790.

\item{[\BBBV]}
C. H. Bennett, E. Bernstein, G. Brassard \& U. Vazirani,
``Strengths and weaknesses of quantum computing'',
\SIAMJC\ {\bf 26} (1997) 1510--1523.

\item{[\BBHT]}
M. Boyer, G. Brassard, P. H{\o}yer \& A. Tapp,
``Tight bounds on quantum searching'',
\ForP\ {\bf 46} (1998) 493--506.

\item{[\Zalka]}
C. Zalka,
``Grover's quantum searching algorithm is optimal'',
\PRA\ {\bf 60} (1999) 2746--2751.

\item{[\Lloyd]}
S. Lloyd,
``Quantum search without entanglement'',
\PRA\ {\bf 61} (1999) 010301.

\item{[\Jozsa]}
R. Jozsa,
``Entanglement and quantum computation'',
in
S. A. Huggett, L. J. Mason, K. P. Tod, S. T. Tsou \&
N. M. J. Woodhouse, eds.,
{\sl The Geometric Universe:  Science, Geometry, and the Work of 
Roger Penrose\/}
(Oxford:  Oxford University Press 1998) 369--379;\hfb
A. Ekert \& R. Jozsa,
``Quantum algorithms:  entanglement-enhanced information 
  processing'',
\PTRSLA\ {\bf 356} (1998) 1769--1782.

\item{[\Steane]}
A. M. Steane,
``A quantum computer needs only one universe'',
{\tt quant-ph/0003084}.

\item{[\NMRseparable]}
S. L. Braunstein, C. M. Caves, R. Jozsa, N. Linden, S. Popescu \&
R. Schack,
``Separability of very noisy mixed states and implications for NMR
  quantum computing'',
\PRL\ {\bf 83} (1999) 1054--1057;\hfb
R. Schack \& C. M. Caves,
``Classical model for bulk-ensemble NMR quantum computation'',
\PRA\ {\bf 60} (1999) 4354--4362;\hfb
For a response to this criticism, see
R. Laflamme,
``Review of `Separability of very noisy mixed states and 
  implications for NMR quantum computing'\thinspace'',
{\sl Quick Reviews in Quantum Computation and Information},
{\tt http://quickreviews.org/qc/}.

\item{[\separabledef]}
R. F. Werner,
``Quantum states with Einstein-Podolsky-Rosen correlations 
  admitting a hidden-variable model'',
\PRA\ {\bf 40} (1989) 4277--4281.

\item{[\Krauss]}
L. M. Krauss,
{\sl The Physics of Star Trek}, 
with a forword by Stephen Hawking
(New York:  HarperCollins 1995).

\item{[\Dirac]}
P. A. M. Dirac,
{\sl The Principles of Quantum Mechanics}, fourth edition
(Oxford:  Oxford University Press 1958).

\item{[\EWL]}
J. Eisert, M. Wilkens \& M. Lewenstein,
``Quantum games and quantum strategies'',
\PRL\ {\bf 83} (1999) 3077--3080.

\item{[\BenjaminHayden]}
S. C. Benjamin \& P. M. Hayden,
``Comment on `Quantum games and quantum strategies'\thinspace'',
{\tt quant-ph/0003036}.

\item{[\SteanevanDam]}
\steane\ \& W. van Dam,
``Physicists triumph at guess my number'',
\PhT\ {\bf 53}(2) (February 2000) 35--39.

\item{[\CEMM]}
R. Cleve, A. Ekert, C. Macchiavello \& M. Mosca,
``Quantum algorithms revisited'',
\PRSLA\ {\bf 454} (1998) 339--354.

\item{[\BernsteinVazirani]}
E. Bernstein \& U. Vazirani,
``Quantum complexity theory'',
in 
{\sl Proceedings of the 25th Annual ACM Symposium on the Theory of
Computing}, San Diego, CA, 16--18 May 1993
(New York: ACM 1993) 11--20;\hfb
This algorithm was rediscovered independently in
B. M. Terhal \& J. A. Smolin,
``Single quantum querying of a database'',
\PRA\ {\bf 58} (1998) 1822--1826.

\item{[\Schrodinger]}
\schrodinger,
``{\it Die gegenw\"artige Situation in der Quantenmechanik\/}'',
\Naw\ {\bf 23} (1935) 807--812; 823--828; 844--849.

\item{[\Schumacher]}
B. Schumacher,
``Quantum coding (information theory)'',
\PRA\ {\bf 51} (1995) 2738--2747.

\item{[\MeyerWallach]}
\dajm\ \& N. R. Wallach,
``Invariants for multiple qubits I:  the case of 3 qubits'',
in preparation.

\item{[\vsmodel]}
G. Salton \& M. McGill,
{\sl Introduction to Modern Information Retrieval\/}
(New York:  McGraw-Hill 1983);\hfb
M. W. Berry, Z. Drma\v c \& E. R. Jessup,
``Matrices, vector spaces, and information retrieval'',
\SIAMR\ {\bf 41} (1999) 335--362.

\item{[\noe]}
\dajm,
``Sophisticated quantum search without entanglement'',
UCSD preprint (2000).

\bye